# Single-Crystal and Powder Neutron Diffraction Study of $Fe_XMn_{1-X}S$ Solid Solutions

Galina Abramova[1], Juerg Schefer[2], Nadir Aliouane[2], Martin Boehm[3], German Petrakovskiy[1], Alexandr Vorotynov[1], Mikhail Gorev[1,6], and Vladimir Sokolov[4]

[1]*Kirensky Institute of Physics, Russian Academy of Sciences, Siberian Branch, Krasnoyarsk, 660036 Russia*
[2]*Paul Scherrer Institute, Laboratory for Neutron Scattering (LNS), CH-5232 Villigen PSI, Switzerland*
[3]*Institut Max von Laue - Paul Langevin, Grenoble Cedex 9, France*
[4]*Nikolaev Institute of Inorganic Chemistry, Siberian Branch, Russian Academy of Sciences, Novosibirsk, 630090 Russia*
[6]*Institute of Engineering Physics and Radio Electronics, Siberian Federal University, prosp. Svobodny 79, Krasnoyarsk, 660041 Russia*

*$Fe_XMn_{1-X}S$ (0 <x< 0.3) synthesized on the basis of α-MnS are the novel Mott-type substances with the rock salt structure. Neutron diffraction study shows that the shift of the Neel temperature of these materials from 150 (x = 0) to 200 K (x = 0.29) under the action of chemical pressure (x) is accompanied by a decrease in the cubic NaCl lattice parameters. The structural transition with a change in crystal symmetry and a decrease in resistance before the magnetic transition at x = 0.25 is found. Therefore, $Fe_XMn_{1-X}S$ are interesting materials for both fundamental study of the interrelation between the magnetic, electrical, and structural properties in systems with the strong electron correlations of a MnO-type and application.*

PACS number_s_: 75.25._z, 61.05._a, 75.80._q, 64.70.Nd

## I. INTRODUCTION

α-MnS belongs to MnO-type substances with the strong electron correlations and the rock salt structure [1,2]. Hydrostatic pressure applied to such systems causes a number of phenomena, including an insulator-to-metal transition, reduction of a magnetic moment (high-spin-to-low-spin transition), and a change in crystal symmetry (ref. in [3,4]). Below Neel temperature $T_N$, the MnO-type oxides and MnS exhibit the FCC II (AFM-II) antiferromagnetic order[5,6]. Under ambient conditions, they are magnetic semiconductors. Under a hydrostatic pressure of 50-100 GPa at room temperature, the MnO-type compounds undergo the insulator-to-metal transition[4]. The effect of the pressure on the magnetic and crystal structures of the compounds still remains understudied [7-9].

In this study, we used the chemical pressure to mimic the changes in physical properties observed by the hydrostatic one in MnS by substituting $Mn^{2+}$ (an ionic radius of 0.97 Å, the high-spin octahedral state) by $Fe^{2+}$ (an ionic radius of 0.92 Å, the high-spin octahedral state). We synthesized new Mott-type compounds with a chemical formula $Fe_XMn_{1-X}S$ (0 ≤ X < 0.3) and the rock salt structure. We report the results of single-crystal and powder neutron diffraction for the $Fe_XMn_{1-X}S$ (0 < X < 0.3) samples and the electrical and structural properties of the latter.

## II. EXPERIMENTAL

Previously, we reported the details of synthesis of $Fe_XMn_{1-X}S$ single crystals and X-ray diffraction data for these materials[10]. Powder neutron diffraction studies of the $Fe_XMn_{1-X}S$ samples with x = 0.05, 0.18, and 0.27 were carried out at the Institut Laue-Langevin (ILL) using a D1A thermal diffractometer with λ = 1.91°A at temperatures of 2-300 K. Single-crystal neutron diffraction studies of the samples with x = 0.25 and 0.29 were carried out at λ = 1.178 Å in a four-circle Eulerian cradle at temperatures of 2-300 K using a TriCS thermal instrument[11] and a SINQ neutron spallation source at the Paul Scherrer Institute (PSI). Simultaneously, we measured the integral intensity and (hkl) position of the magnetic and nuclear peaks for the $Fe_XMn_{1-X}S$ samples. The neutron and X-ray diffraction data obtained were refined using a Fullprof program[12] based on the Rietveld method. Resistance and lattice strain were measured at the Kirensky Institute of Physics. The electrical properties were investigated with the use of a PPMS facility (Quantum Design) at temperatures of 2-300 K. Thermal lattice expansion measurements were performed in the temperature range 120–300 K with a heating rate of 3 K min$^{-1}$ using a NETZSCH DIL-402C push-rod dilatometer with a fused silica sample holder. The results were calibrated by taking $SiO_2$ as a reference to eliminate the influence of thermal expansion of the system. Parallelepiped samples were cut from large single crystals so as to have parallel surfaces perpendicular to the (100) crystallographic directions; distance L between the MnS surfaces was 4.915 mm. These samples were used in electrical and thermal lattice expansion measurements.

## III. RESULTS AND DISCUSSION

### A. Neel temperature of $Fe_XMn_{1-X}S$

X-ray study showed that, at room temperature, the $Fe_XMn_{1-X}S$ single crystals, similar to the α-MnS phase, are characterized by the face-centered cubic (FCC) structure of a NaCl type (Fm-3m space group). Substitution of iron ions $Fe^{2+}$ for manganese ions $Mn^{2+}$ in $Fe_XMn_{1-X}S$ leads to a decrease in the lattice parameter from 5.224 (x = 0) to 5.165 Å (x = 0.29). The obtained parameter of a FCC unit cell at 300 K (Fig.1) for the $Fe_XMn_{1-X}S$ samples with x = 0.29 (under the action of chemical pressure $x$) is close to the value observed in MnS under a hydrostatic pressure of 3–4 GPa [13]. A decrease in the cubic lattice parameter under the chemical pressure in $Fe_XMn_{1-X}S$ at room temperature is accompanied by reduction of electrical resistance by six orders of magnitude. We found that, at 2-300 K, conductivity of the $Fe_XMn_{1-X}S$ samples changes from the semiconductor type for x < 0.25 to metal one for x ≥ 0.25. This indicates the strong correlation between the electrical properties of $Fe_XMn_{1-X}S$ and the interatomic distance of the magnetic ions.

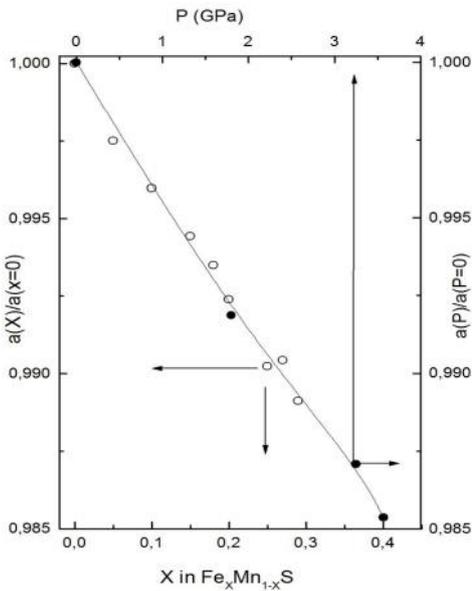

FIG.1. The change of relative parameter of the cubic NaCl lattice under hydrostatic pressure (P) for MnS [13] and upon Fe substitution (x in $Fe_XMn_{1-X}S$) at room temperature.

Single-crystal and powder neutron diffraction data taken at different temperatures on a D1A (ILL), TriCS, and HRPT (PSI) showed that the structure of the $Fe_XMn_{1-X}S$ samples is similar to that of MnS. The single-crystal and powder neutron diffraction patterns above the Néel transition contain only odd (111, 113, 331, and 511) Bragg reflections corresponding to the nuclear reflections, which are typical of the MnS paramagnetic state (Fm-3m space group). Substitution of Fe ions in the $Fe_XMn_{1-X}S$ samples by Mn ions is accompanied by a decrease in the integral intensity of the nuclear Bragg reflections from the odd planes and an increase in the integral intensity of the nuclear Bragg reflections from the even planes, because manganese (-0.37) and iron (+0.96) ions have negative and positive neutron scattering amplitudes (in the units of $10^{-12}$ cm [14-15]). Ions of Fe and S have positive neutron scattering amplitudes. For this reason, the neutron pattern of FeO differs from the patterns of MnS and MnO by the presence of the even planes [5, 6, 16]. Our $Fe_XMn_{1-X}S$ compounds exhibit the varying integral intensity of the reflections on the patterns upon replacement of $Mn^{2+}$ ions (S = 5/2) by $Fe^{2+}$ ions (S = 2). These data also confirm the formation of the $Fe_XMn_{1-X}S$ solid solutions.

Below Neel temperature $T_N$, MnO, FeO, and MnS exhibit the FCC II (AFM II) antiferromagnetic order with ferromagnetic (FM) sheets of the (111) planes that are antiferromagnetically stacked along the [111] direction [5,6,16]. The lattices of MnO, FeO, and NiO are trigonaly distorted along the [111] direction; thus, the lattice symmetry is consistent with the symmetry of the magnetic structure [5,6]. The magnetic reflexes caused by the AFM-II order appear at the (h/2 k/2 l/2) reciprocal lattice points, where h, k, and l are the odd numbers.

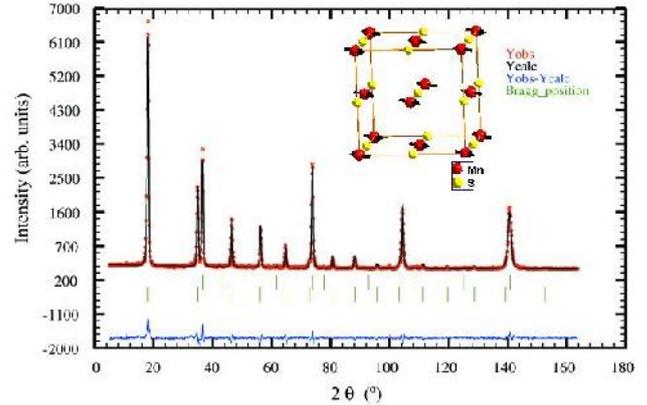

FIG.2. Powder neutron diffraction patterns of MnS at 20 K. Insert: direction of the magnetic moments of Mn ions in the NaCl lattice.

Figure 2 presents powder neutron diffraction patterns of MnS that were taken on a HRPT (λ = 1.886 A) at 20 K. The magnetic structure of the MnS sample is similar to that reported in [16]. The direction of the magnetic moments of Mn ions in the MnS sample was obtained to corresponds to <110> (insert in Fig.2); the magnetic moments are ferromagnetically ordered on the (111) planes and antiferromagnetically ordered between these planes with the spin propagation vector (1/2,1/2,1/2). Figure 3 presents the powder neutron diffraction patterns of the $Fe_XMn_{1-X}S$ samples taken on a D1A (ILL) (λ = 1.911 A) at 2 K. The magnetic structure of $Fe_XMn_{1-X}S$ at x ≈ 0.1 is typical of A-type antiferromagnets with the Roth classification [6]. The magnetic structure of $Fe_XMn_{1-X}S$ at x ≈ 0.2 can be presented as a B-type antiferromagnet, used the Roth classification [6]. The observed magnetic Bragg reflections are of the odd type only. The intensity of all odd magnetic peaks with spin propagation vector (1/2,1/2,1/2) vanishes at $T_N$.

Figure 4a shows temperature dependences of the relative integral intensity of the magnetic Bragg reflections I(T)/I(2 K) observed for the $Fe_XMn_{1-X}S$ samples. The integral intensity of one peak on the pattern is given by $I = \sqrt{\pi}$ (Imax/FWHM), where FWHM is the total width at a half maximum of the peak. Magnetization M of the magnetic sublattice is proportional to the square root of the measured neutron intensity (I). In $Fe_XMn_{1-X}S$, we found a shift of the Neel temperature from 150 (x = 0) to 200±3 K (x = 0.29) related to an increase in the chemical pressure (X). This indicates the enhancement of the super-exchange interaction constant with decreasing distance between magnetic atoms upon substitution Mn→Fe. The estimated values of the exchange integral of the superexchange interaction along the <100> direction for the (Mn,Fe)- S - (Mn,Fe) bonds are $J^{180}$ = - 4.15 K for MnS and $J^{180}$ = - 7.27 K for $Fe_XMn_{1-X}S$ (x = 0.29). Thus, substitution Mn→Fe causes a decrease in cubic lattice parameter *a* and an increase in the Neel temperature. The Neel temperature shifts occurring in MnS upon this substitution for our case and under hydrostatic pressure ([17]) are compared in Fig. 5. A linear increase of the Neel temperature with increasing X in $Fe_XMn_{1-X}S$ is observed only in the case of semiconductor substances (x < 0.25).

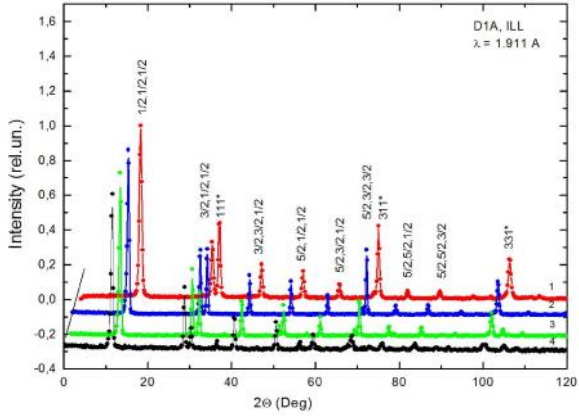

FIG.3. Powder neutron diffraction patterns of $Fe_XMn_{1-X}S$ with x = 0 (1);=0.05(2);=0.18(3);=0.27(4) at 2 K.

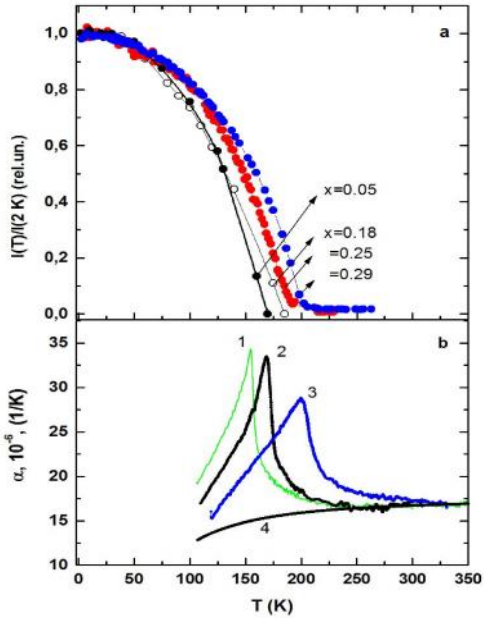

FIG. 4. Temperature dependences of the relative integral intensity of the magnetic reflections (a) and thermal expansion coefficient α (b) for $Fe_XMn_{1-X}S$ samples, b: MnS (1); x=0.1 (2); x=0.29 (3); without the anomalous lattice behavior (4).

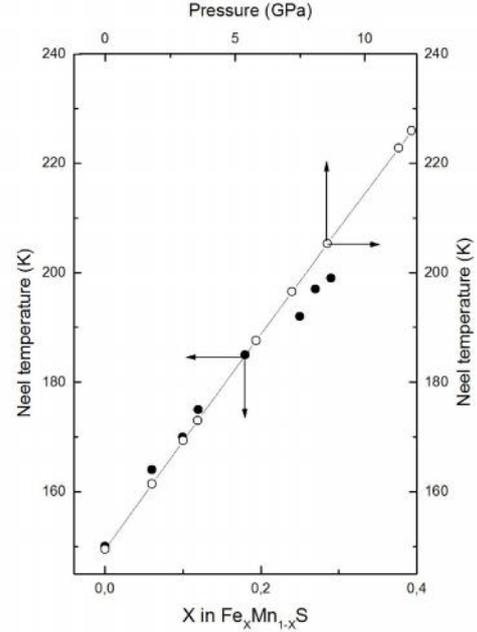

FIG.5. Comparison of the Neel temperature shift in MnS under hydrostatic pressure [13] and upon Fe substitution (our investigation).

### B. Lattice strain in $Fe_XMn_{1-X}S$

Small rhombohedral distortion of the cubic NaCl lattice of MnS was found by B. Morosin [18] at a Neel temperature of 150 K and by H. van der Heide et al. [19] at about 200 K. We found no substantial evidences for the change of symmetry lattice of $Fe_XMn_{1-X}S$ with 0 < x < 0.25 at low temperatures (2–7 K). However, as can be seen in Fig. 4b, there is an anomaly of thermal expansion coefficient α of these samples at the Neel temperature (α is measured along the [100] lattice direction, i. e., for the (100) planes). This figure shows temperature dependences of the thermal lattice expansion coefficient α = (1/L)dL/dT of the $Fe_XMn_{1-X}S$ single crystals. The anomalous contribution to α is observed in a wide temperature range (for example, 100-260 K for MnS), being maximum at the Neel temperature.

The temperature dependence of magnetization $M_{rel}$ (in relative units) obtained by Imax(T)*FWHM (T = 100 K)/Imax(T = 100 K)*FWHM(T) for $Fe_XMn_{1-X}S$ (x = 0.05, $T_N$ = 165 K) is shown in Fig. 6. For comparison, in the same figure the temperature dependence of relative anomalous contribution $S_{rel}$ to the lattice strain of MnS ($T_N$ = 150 K) is shown. The relative value of the anomalous lattice strain is obtained by $S_{rel}$ = $dL/L_m$ (T)/$dL/L_m$ (T=100 K), $S_{rel}$ < 0 (not shown in the figure). The temperature dependences of the structural and magnetic properties are consistent, which allows us to conclude that the anomalous contribution $dL/L_m$ < 0 to the lattice strain near the Neel temperature is due to the magnetic transition (the negative magnetoelastic effect along the [100] lattice direction for MnS). Magnetoelastic constants *B1* along the [100] direction and *B2* along [111] direction of the FCC lattice can be estimated from the anomalous contribution to the lattice strain $(dL/L)_{m<100>}$ = - (2/3) B1 /($C_{11}$-$C_{12}$) and $(dL/L)_{m<111>}$ = - (2/3) B2 /$C_{44}$ in the magnetically ordered state, $C_{11}$, $C_{44}$, and $C_{12}$[17] are the elastic modules for MnS.

The observed value of $(dL/L)_m$ along [100] is -0.0008 at 106 K for MnS, $(C_{11}-C_{12}) = 0.69 \cdot 10^{+12}$ din/cm$^2$. The estimated value B1 = 83 J/cm$^3$ for MnS at 106 K is typical of the spinel compounds of 3d-elements.

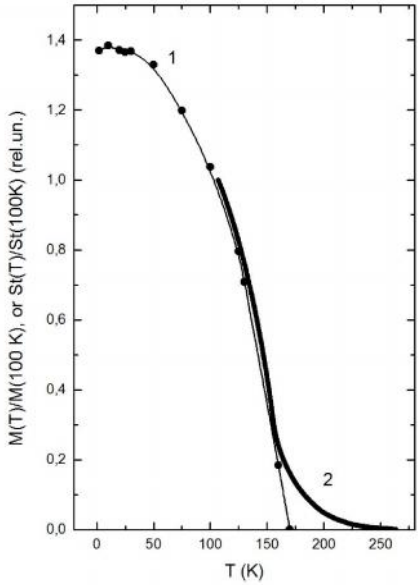

FIG. 6. Comparison of relative magnetization $M_{rel}$ of $Fe_XMn_{1-X}S$ with x =0.05 (curve 1) and relative anomalous lattice strain $S_{rel}$ of MnS (curve 2, negative anomalous lattice strain).

### C. Properties of the single crystals $Fe_XMn_{1-X}S$ with $0.25 \leq x \leq 0.29$

The step-wise semiconductor-to-metal electron transition in pure MnS is observed at room temperature under a pressure of about 26 GPa [20]. Our investigation of the electrical resistance shows that MnS is the semiconductor in the temperature range of 77-300 K and it undergo the change of the conductivity type under the chemical pressure. The temperature dependence of electrical resistance of the $Fe_XMn_{1-X}S$ at x = 0.25 is shown in Fig. 7a, for the illustration. The presented data clearly indicate metal conductivity; the resistance decreases with temperature below 240-250 K. The magnetic properties of single crystals of $Fe_XMn_{1-X}S$ at x = 0.25 and 0.29 were investigated by a TriCS method at the PSI. The temperature dependences of the integral intensity of the magnetic peaks (Fig. 7c) are similar to those of the $Fe_XMn_{1-X}S$ samples with x < 0.25. The temperature hysteresis of magnetization of about 10° at the Neel temperature is found upon heating and cooling the samples with x = 0.25 and 0.29. This fact indicates the occurrence of the first-order magnetic transition. Small residual intensity of the magnetic peaks is observed at $T > T_N$ in zero magnetic field and may indicate the occurrence of the short-range order up to 220 K. We have found the $Fe_XMn_{1-X}S$ single crystals with $x \geq 0.25$ undergo a structural phase transition with a change of the crystal symmetry additional to the magnetic transition. Figure 7b presents temperature dependences of the unit cell parameters for the sample with x = 0.25, which are calculated from the nuclear reflection of (111) and (311). The magnetic structure can be preliminary described by three propagation vectors: k1 = (1/2,1/2,1/2,), k2 = (1/2,1/2,0), and k3 = (1/2,0,0).

However, to make a correct conclusion, one should clarify the symmetry of a new nuclear lattice of these samples. Temperature evolution of the (111) reflection d-spacing and lattice parameter a(111) of NaCl phase for the sample with x = 0.25 (Fig.7b) is similar to MnO, which is obtained by the authors [21] at 3.3 GPa. Such behavior the (111) reflection can be explain by splitting of this reflection into two reflections (003 and 101) due to the structural transition from the cubic Fm-3m to rhombohedral R3m – like phase. The geometry of our neutron experiment alloy to us the observation only the transformation of the (111) reflection of Fm-3m state to (003) reflection of R3m phase.

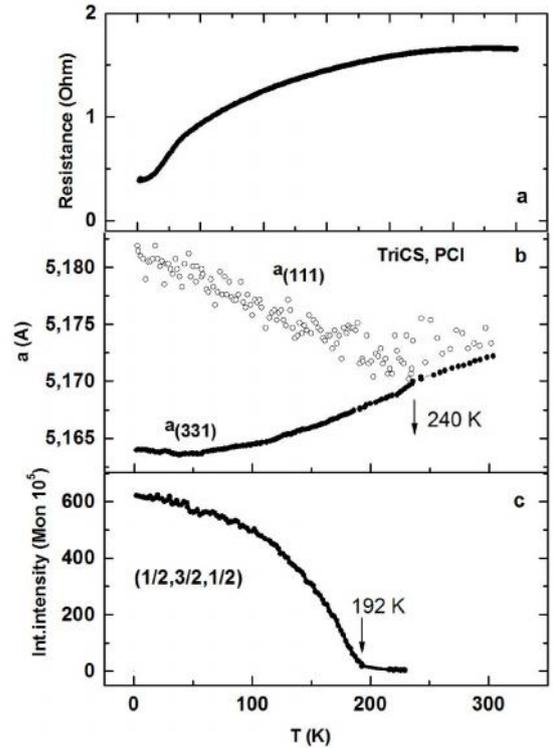

FIG. 7. Temperature dependences of electrical resistance (a), lattice parameter (b), and integral intensity of the magnetic peak (1/2;3/2;1/2) (c) for x = 0.25.

It can be seen from Fig.7b that the structural transition is observed at about 240 K and precedes the magnetic transition $T_N$ = 192 K. This structural transition is accompanied by the change in the conductivity type from the high-temperature semimetal state to the low-temperature metal one (Fig. 7a) upon cooling the sample. Since the critical temperatures of the structural transformation and of the magnetic transition in $Fe_XMn_{1-X}S$ are different, the mechanism of the observed structure transition may be of nonmagnetic origin. The structural transition and the change in the conductivity type break a linear increase in the Neel temperature of the $Fe_XMn_{1-X}S$ solid solutions (Fig. 5a). At the same time, the temperature range of the local short-range magnetic order in the $Fe_XMn_{1-X}S$ samples 0.25 < x<0.29 extends up to 300 K, especially, in the applied magnetic field. The more detailed data of the $Fe_XMn_{1-X}S$ samples 0.25 < x<0.29 will be presented in next manuscript.

## IV. CONCLUSIONS

We studied the effect of the chemical pressure (Fe cation substitution) on the magnetic, electrical, and structural properties of α-MnS belonging to the MnO-type substances at temperatures of 2-300 K. We found that the chemical pressure leads to a decrease in the cubic lattice parameter and in electrical resistance and causes the insulator-to-metal transition. A linear increase in the Neel temperature is observed by the neutron diffraction for the semiconductor substances $Fe_XMn_{1-X}S$ ($0 < x < 0.25$) with increasing Fe cation substitution. The magnetic structure of $Fe_XMn_{1-X}S$ solid solutions at $x \approx 0.1$ as in MnS is typical of A-type antiferromagnets. The magnetic structure of $Fe_XMn_{1-X}S$ at $x \approx 0.2$ can be presented as a B-type antiferromagnets, used the Roth classification. For the metal substances $Fe_XMn_{1-X}S$ ($0.25 < x < 0.29$), the change in the nuclear and magnetic lattice symmetry was found.

## V. ACKNOWLEDGMENTS

We acknowledge the support of the Institut Laue-Langevin (Grenoble, France), the spallation neutron source SINQ (TriCS, HRPT) and the SLS-MS beamline, and the Paul Scherrer Institute (Villigen, Switzerland). We thank the researches of the Shared Facilities Center of the Krasnoyarsk Scientific Center (Krasnoyarsk, Russia) for help in the experiments.